\documentclass[aps,prb,twocolumn,superscriptaddress, longbibliography]{revtex4-2}

\usepackage{graphicx}
\usepackage{dcolumn}
\usepackage{bm}
\usepackage{hyperref}
\usepackage{todonotes}
\usepackage{color}
\usepackage{glossaries}
\usepackage[percent]{overpic}

\def\um{\text{µ}\text{m}}

\begin{document}

\preprint{APS/123-QED}

\title{Stacking disorder in $\alpha$-RuCl$_3$ via x-ray three-dimensional difference pair distribution function analysis}

\author{J. Sears}
\author{Y. Shen}

\affiliation{Department of Condensed Matter Physics and Materials Science, Brookhaven National Laboratory, Upton, New York 11973, USA}

\author{M. J. Krogstad}
\affiliation{Advanced Photon Source, Argonne National Laboratory,
Lemont, Illinois 60439, USA}

\author{H. Miao}
\author{Jiaqiang Yan}
\affiliation{Material Science and Technology Division, Oak Ridge National Laboratory, Oak Ridge, Tennessee 37830, USA}

\author{Subin Kim}
\affiliation{Department of Physics, University of Toronto, 60 St. George Street, Toronto, Ontario M5S 1A7, Canada}

\author{W. He}
\author{E. S. Bozin}
\affiliation{Department of Condensed Matter Physics and Materials Science, Brookhaven National Laboratory, Upton, New York 11973, USA}

\author{I. K. Robinson}
\affiliation{Department of Condensed Matter Physics and Materials Science, Brookhaven National Laboratory, Upton, New York 11973, USA}
\affiliation{London Centre for Nanotechnology, University College,
Gower St., London WC1E 6BT, UK}

\author{R. Osborn}
\author{S. Rosenkranz}
\affiliation{Materials Science Division, Argonne National Laboratory, Lemont, IL 60439, USA}

\author{Young-June Kim}
\affiliation{Department of Physics, University of Toronto, 60 St.\ George Street, Toronto, Ontario M5S 1A7, Canada}

\author{M. P. M. Dean}
\email{mdean@bnl.gov}
\affiliation{Department of Condensed Matter Physics and Materials Science, Brookhaven National Laboratory, Upton, New York 11973, USA}

\date{\today}

\begin{abstract}
The van der Waals layered magnet $\alpha$-RuCl$_3$ offers tantalizing prospects for the realization of Majorana quasiparticles. Efforts to understand this are, however, hampered by inconsistent magnetic and thermal transport properties likely coming from the formation of structural disorder during crystal growth, postgrowth processing, or upon cooling through the first order structural transition. Here, we investigate structural disorder in $\alpha$-RuCl$_3$ using x-ray diffuse scattering and three-dimensional difference pair distribution function (3D-$\Delta$PDF) analysis. We develop a quantitative model that describes disorder in $\alpha$-RuCl$_3$ in terms of rotational twinning and intermixing of the high and low-temperature structural layer stacking. This disorder may be important to consider when investigating the detailed magnetic and electronic properties of this widely studied material. 

\end{abstract}

\maketitle

\section{Introduction}

The exactly solvable Kitaev model on the two-dimensional honeycomb lattice is a fascinating example of how frustrated magnetic interactions can lead to the emergence of novel fractionalized quasiparticles such as Majorana fermions and gauge fluxes \cite{kitaev2006}. Due to this, the identification of the candidate Kitaev material $\alpha$-RuCl$_3$ attracted considerable attention \cite{plumb2014, banerjee2016, kasahara2018majorana, trebst2022}. RuCl$_3$ comprises honeycomb layers of Ru atoms coordinated with Cl which are stacked via weak van der Waals bonds to form bulk crystals. Although an ideal realization of the Kitaev model would be a purely two-dimensional system that remains a quantum spin liquid down to zero temperature, RuCl$_3$ exhibits three-dimensional magnetic order. 

Recent years have seen extensive efforts to better understand the properties of RuCl$_3$. Early reports observed that the RuCl$_3$ magnetic ordering temperature varied substantially from sample-to-sample \cite{banerjee2016}. Later, thermal Hall measurements, which are of particular importance in identifying the presence of chiral Majorana edge modes, also suffered similar repeatability challenges
\cite{kasahara2018majorana, kasahara2022, czajka2021oscillations, czajka2023planar}. A natural reason for these inconsistencies would be RuCl$_3$'s propensity towards stacking disorder which is associated with the combination of weak van der Waals interlayer bonding and the different near-degenerate forms of inter-layer stacking that tend to exist in layered honeycomb van der Waals materials \cite{hskim2016, mcguire2015, mcguire2017}. These types of structural disorder are often considered to be at the root of the puzzling sample-dependent properties of RuCl$_3$ \cite{banerjee2016, zhang2023sampledependent, zhang2023stacking}.

The accepted room-temperature RuCl$_3$ structure has the $C2/m$ (\#12) space group with $a=5.98$~\AA{}, $b=10.34$~\AA{}, $c=6.01$~\AA{}, $\alpha=\gamma=90^{\circ}$, and $\beta=108.9^{\circ}$ \cite{johnston2015}. As shown in Fig.~\ref{fig1}(a), this involves neighboring planes being displaced along the armchair direction of the layers, as illustrated by the green vector. More significant sample variability has been observed at low temperatures, below a widely reported and highly hysteric structural transition which generally occurs within the temperature range of 50 to $170$~K \cite{park2016, cao2016, li2021giant, lebert2022, mu2022}. At low temperature the structure is believed to be quite disordered, but a significant $R\bar{3}$ (\#148) structural component is thought to emerge \cite{park2016}. This involves the Ru atoms in one layer sitting over the center of the hexagon from the layer below [Fig.~\ref{fig1}(a)].

A comprehensive understanding of the intriguing transport properties of RuCl$_3$ will likely require an evaluation of the local structural properties, including stacking order and disorder. The pair distribution function (PDF) method is often a useful tool for such a problem, but in this case preparing unstrained and non-textured powder samples presents a major challenge. 3D-$\Delta$PDF methods are compatible with single crystals and yield richer structural information \cite{Weber2012three, Krogstad2020reciprocal}, which extends what can be learned from diffuse scattering \cite{Hendricks1942x, Welberry2016one}. Here we present a 3D-$\Delta$PDF analysis of the x-ray diffuse scattering from a single crystal sample of RuCl$_3$. We find that, within our resolution of 0.3~\AA{}, the diffuse scattering has the form expected for a crystal composed of identical layers with the disorder arises from stacking faults. Using 3D-$\Delta$PDF analysis we provide a comprehensive overview of the types of layer stacking and estimate the prevalence of the different types of stacking faults. Our measurement also allows us to place an upper constraint of 0.3~\AA{} on the size of any distortions that may occur within the honeycomb layers that occur as a result of the layer restacking. 

\section{Methods} \label{method}

\begin{figure*}[!hbtp]
\includegraphics[width=2\columnwidth]{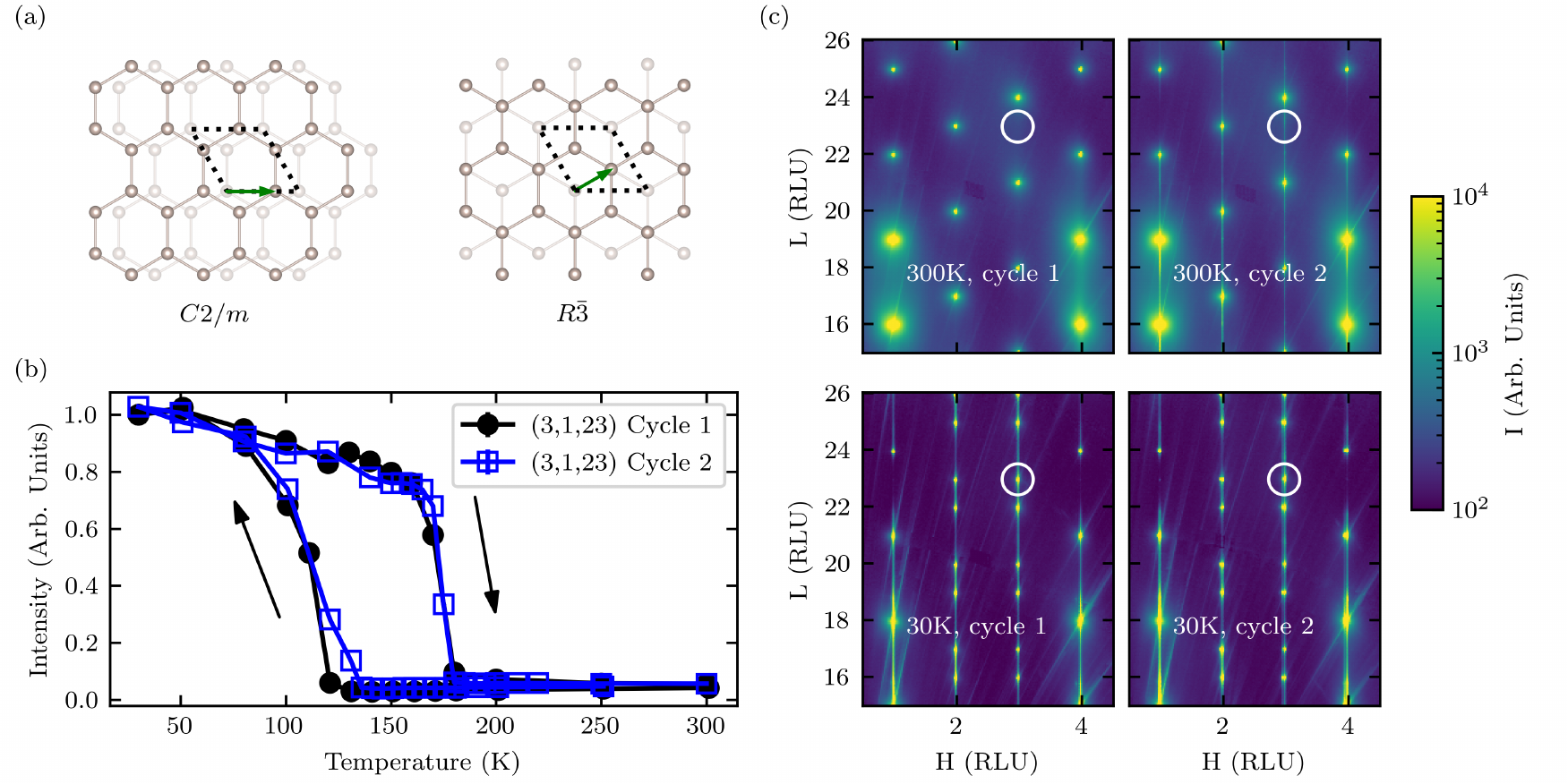}
\caption{Characterization of the structural phase transition. (a) Layer stacking in the high temperature $C2/m$ structure, and in the  $R\bar{3}$-like structure that is prevalent at low temperatures. The dotted outline shows the hexagonal crystallographic supercell described in the methods. The figure shows only the ruthenium honeycomb layers, omitting chlorine atoms for clarity.  (b) Integrated intensity of the (3,1,23) Bragg peak normalized to the intensity at 30~K. This peak is present only in the low temperature structural phase. The structural transition is observed to have a large hysteresis of $\sim$50~K. (c) $H$-$L$ reciprocal space maps at $K$=1, for a single crystal at 300~K and 30~K. The top left panel shows the data from the as-grown crystal, that had never been cooled to low temperature. We observed very little diffuse scattering in this data, indicating that the crystal was of good quality with very little disorder. The white circles indicate the Bragg peak plotted in panel (b). These maps show the changes in Bragg peak positions due to the structural transition, as well as the appearance of rods of weak, diffuse scattering which remain even after rewarming the sample to room temperature. The left two panels labeled `cycle 1' were collected on the first cooling of the sample, while the right two panels (`cycle 2') were collected on the second cooling. Intensity is plotted on a logarithmic scale to enhance visibility of the weak, diffuse features. }
\label{fig1}
\end{figure*}

Throughout this work, we will index all peaks in terms of a simple hexagonal supercell with lattice parameters $a=b=5.98$~\AA{}, $c=17.18$~\AA{}, $\alpha=\beta=90^\circ$, $\gamma=120^\circ$. This unit cell, which is shown as the black dotted lines in Fig.~\ref{fig1}(a), is chosen for convenience as it allows all Bragg peaks to be indexed to integer values, and because its $ab$ plane coincides with the honeycomb plane. The $c$ lattice parameter of this supercell is equal to three times the distance between honeycomb layers.

Single crystals of $\alpha$-RuCl$_3$ were grown at the Oak Ridge National Laboratory using chemical vapor transport technique in a two-zone tube furnace with the hot end kept at 1000$^{\circ}$C and the cold end at 750$^{\circ}$C. RuCl$_3$ powder obtained by reacting RuO$_2$ with AlCl$_3$-KCl salt mixture \cite{yan2017} was used for the crystal growth. Flat plates with lateral size $\sim$500~\um{} were used in this study. The sample was cooled using a helium cryocooler and measurements were collected from room temperature down to a base temperature of 30~K.

X-ray diffraction data spanning large regions of reciprocal space were collected at the 6-ID-D beamline at the Advanced Photon Source at the Argonne National Laboratory. The X-ray energy was 87 keV to minimize absorption and maximize reciprocal space coverage. The beam size was chosen to over-illuminate the sample. A Pilatus 2M CdTe detector was used to collect the diffraction data. The sample was rotated continuously through a full 360$^\circ$ and the image was read from the detector every 0.1$^\circ$. This measurement was completed at two further distinct axes of rotation to improve coverage of reciprocal space and to cover the gaps between the sensors on the detector. The data obtained covered a range of approximately $\pm$11 \AA{} $^{-1}$. Data were taken with different X-ray attenuator settings in order to obtain data without saturation-related distortions to the Bragg peak intensity and also high signal-to-noise diffuse scattering. The images were processed and pixels binned into a grid of voxels in reciprocal space, to generate a 3D volume of reciprocal space data using approaches developed in Ref.~\cite{Krogstad2020reciprocal}.

\begin{figure*}[!hbtp]
\includegraphics[width=2\columnwidth]{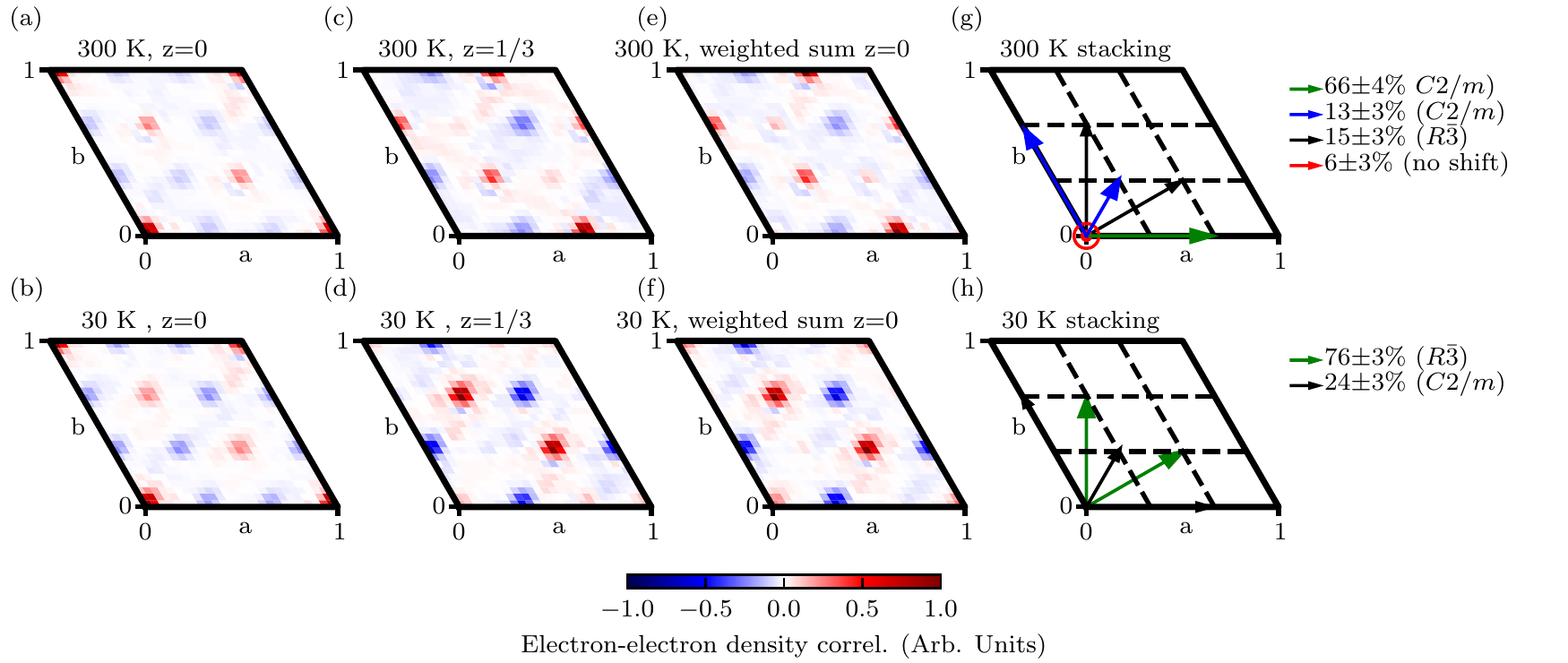}
\caption{
Inter-layer stacking characterization by 3D-$\Delta$PDF. The upper row shows 300~K data (after temperature cycling), while the bottom row shows 30~K. (a), (b) $x-y$ cut of the 3D-$\Delta$PDF at $(x=3, y=3, z=0)$. We note that these cuts are not blank, as would be expected for a 3D-$\Delta$PDF without intralayer disorder. This is due to the stacking faults which have the effect of reducing the in-plane lattice constants by $\frac{1}{3}$ as discussed in the text. As a result, this is a 3D-$\Delta$PDF with respect to the average structure with lattice constants $\frac{a}{3}$. (c), (d) $x-y$ cut of the 3D-$\Delta$PDF at $(x=3, y=3, z=\frac{1}{3})$, where $z=\frac{1}{3}$ corresponds to the distance between adjacent layers. (e), (f) Weighted sums of several copies of the $(x=3, y=3, z=0)$ cut of the 3D-$\Delta$PDF, translated by different in-plane vectors corresponding to the interlayer stacking types, which reproduce the observed $z=\frac{1}{3}$ 3D-$\Delta$PDF in (c),(d), respectively. (g),(h) The translations and weighting factors used to generate the weighted sums shown in (e) and (f) that describe the stacking disorder present in RuCl$_3$. The stacking types for the long range ordered structure are shown as green arrows, while the blue, black and red arrows represent stacking types found only in the disordered component of the crystal. Listed percentage contributions are evenly divided between all the translation vectors of that type. Errorbars were estimated by manually adjusting the weighting factors until the weighted sum differed visibly from the $z=\frac{1}{3}$ data.  }
\label{fig2}
\end{figure*}

To perform the 3D-$\Delta$PDF analysis, it is necessary to separate the Bragg peak and diffuse scattering components of the data. Bragg peak data were isolated by selecting regions of reciprocal space according to the structural selection rules of the nominal structural space group ($C2/m$ at high temperature and $R\bar{3}$ at low temperature). The range of reciprocal space around the Bragg peaks was chosen by observing the change in gradient between the Bragg and diffuse scattering. Checks were performed to test whether the results obtained depend on the punch size. The diffuse scattering data were isolated by ``punching out'' the Bragg peak data and filling in the gaps using a linear interpolation along the reciprocal space $L$ direction. The diffuse scattering data was then Fourier transformed to produce 3D-$\Delta$PDF datasets \cite{Krogstad2020reciprocal}.

\section{Reciprocal space analysis}

We begin by considering the reciprocal space diffraction data on RuCl$_3$ as plotted in Fig.~\ref{fig1}. The transition between the room temperature monoclinic $C2/m$ structure and the low temperature nominally rhombohedral  $R\bar{3}$ phase can be tracked by the appearance of Bragg peaks such as (3,1,23), which only occur in the $R\bar{3}$ phase. This transition may be thought of primarily as a restacking of the honeycomb layers, as shown in Fig.~\ref{fig1}a (chlorine atoms have been omitted for clarity). Figure~\ref{fig1}(b) shows clear signatures of this transition with a large hysteresis. As the sample is cooled through the transition the Bragg peaks specific to the $C2/m$ phase disappear, and new peaks appear at the positions expected for the low temperature $R\bar{3}$ unit cell.

The crystal selected for this measurement was an untwinned crystal, with Bragg peaks well indexed by a single domain of the known $C2/m$ structure at room temperature. Upon temperature cycling into the low temperature structure, which is proposed to be $R\bar{3}$, and back to room temperature we observe a return to the same untwinned $C2/m$ structure. This shows that the crystal ``remembers'' the specific direction defining the $C2/m$ structure even though the $R\bar{3}$ structure would suggest that the crystal is nominally three-fold symmetric. 

In addition to the changes to the long range structure, we also observed the appearance of rods of diffuse scattering upon the first cooling of the crystal. This is shown in the four panels of Fig.~\ref{fig1}c, plotting the $(H1L)$ reciprocal lattice plane at high and low temperature, as the crystal is cycled in temperature. The crystal initially showed only a small amount of diffuse scattering, which manifests as lobes of scattering with a shape typical of thermal diffuse scattering. As such, the as-grown crystal does not show evidence for structural disorder. There is a substantial increase in diffuse scattering upon cooling into the low temperature phase. This new scattering appears as extended rods along the reciprocal space $L$ direction, perpendicular to the honeycomb planes. These rods of scattering remain even when the crystal is warmed back into the high temperature phase. Our observed pattern of diffuse intensity, involving rods of scattering along $L$, is indicative of incorrect stacking, or horizontal translation of individual planes within the crystal structure. The lack of strong structural diffuse scattering between the rods suggests that minimal disorder is present within the individual planes. We will use the hypothesis that disorder in RuCl$_3$ is composed of stacking faults in our more detailed 3D-$\Delta$PDF analysis in the next section. 

\begin{figure*}[!hbtp]
\includegraphics[width=2.\columnwidth]{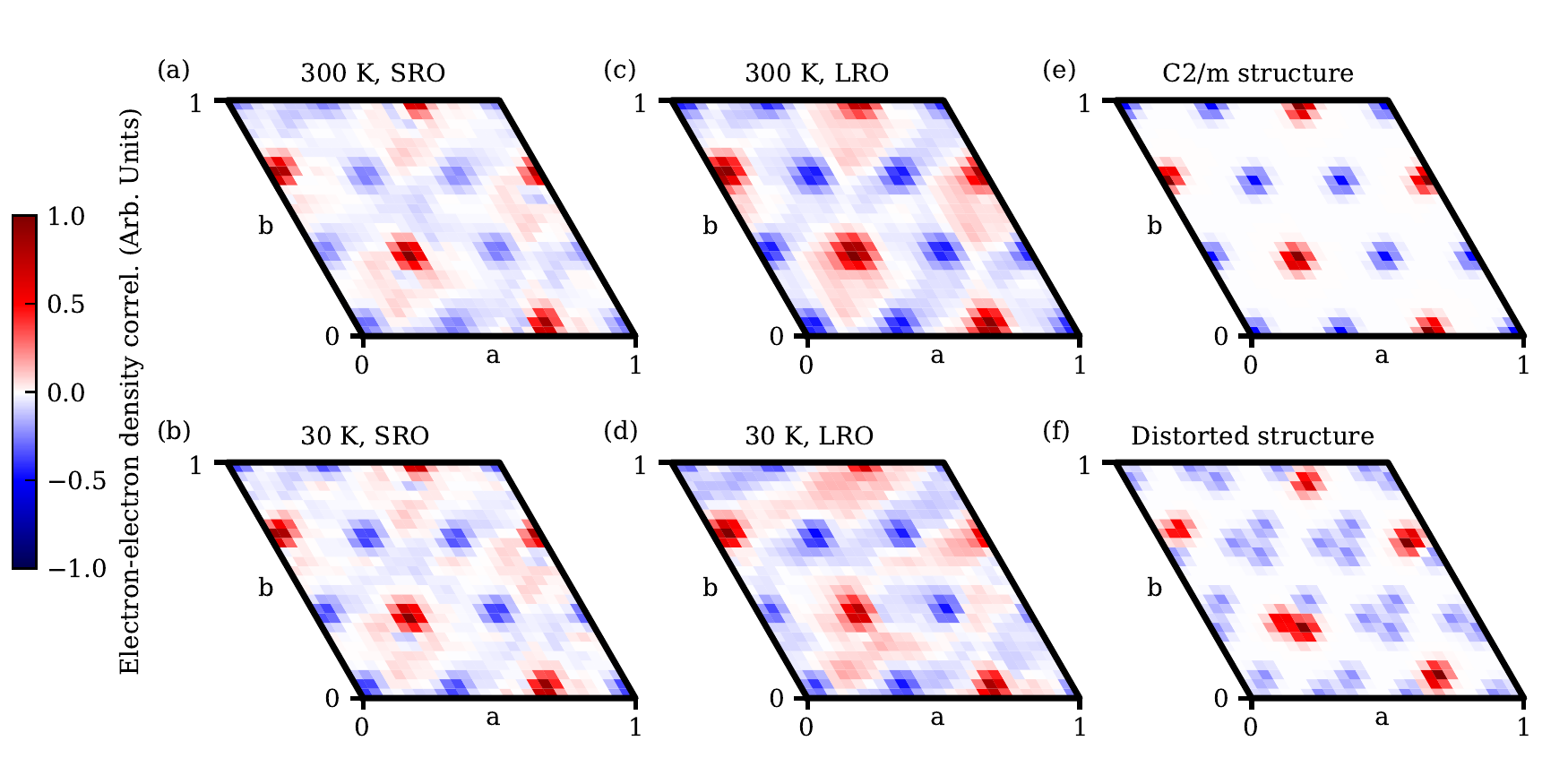}
\caption{
Single layer structural characterization of RuCl$_3$ via real-space correlation functions. (a), (b) 3D-$\Delta$PDF computed from data with the Bragg peaks removed leaving only the diffuse component of the scattering (denoted short range order or SRO). (c), (d) ``pseudo $\Delta$PDF'' that reflects the long range order (LRO). (e) Simulation using the $C2/m$ structural model. (f) Simulation using the highly distorted structure found for single layer exfoliated RuCl$_3$ \cite{Yang2023}. We observe that within the measurement resolution, all single layer structures agree with the monoclinic single layer structure. No distortions of the magnitude seen in the exfoliated layer are observed in the bulk crystal. All data are taken at $(x=6, y=3, z=0.08)$, where $z=0.08$ corresponds to the Ru-Cl distance. We made this choice over the $z=0$ plane in order to have higher sensitivity to the distortions observed in Ref.~\cite{Yang2023}.}
\label{fig3}
\end{figure*}

\section{Stacking disorder}

The stacking faults underlying the observed diffuse scattering were assessed by 3D-$\Delta$PDF analysis. The Bragg peaks were punched out as described in the methods section, isolating the diffuse scattering which was Fourier transformed to produce the 3D-$\Delta$PDF. This 3D-$\Delta$PDF encodes differences in the electron density-density correlation with respect to the average crystal structure \cite{Weber2012three}. It is important to note that the average crystal structure will not necessarily correspond to the nominal unit cell as different types of correlated disorder can affect both the average structure and the local structure. In particular, the average structure in a material with stacking faults comprises a superposition of the possible atomic positions from all different layers. This means that the unit cell describing the average structure contains atoms at all possible locations, but with partial atomic occupancy related to the prevalence of each atomic position. Most simple types of random stacking faults will lead to an equal population of each atomic position. 

We start by inspecting the 3D-$\Delta$PDFs at $z=0$, as shown in Fig.~\ref{fig2}(a)\&(b). One might naively expect that if all disorder arises from stacking faults, with no disorder in the individual planes, then the $z=0$ 3D-$\Delta$PDF should be blank. On the contrary, we observe patterns of positive and negative spots at positions that are combinations of $(\frac{1}{3}, 0)$ and $(0, \frac{1}{3})$ in-plane displacement vectors. These displacements imply nine possible layer positions within the $(x, y)$ plane, which are superimposed in the average structure. This results in an average unit cell with in-plane lattice constants of $\frac{a}{3}$, and resulting partial atomic occupancy. The $z=0$ 3D-$\Delta$PDF, can then be understood as having signal at locations corresponding to Ru-Ru and Cl-Cl interatomic vectors, with either positive or negative density depending on whether the real occupancy is larger or smaller than the partial atomic occupancy in the smaller average unit cell. The $z=0$ 3D-$\Delta$PDF plane thus serves as a basis for understanding RuCl$_3$ overall.

We now consider how to extract information about the stacking faults from the 3D-$\Delta$PDF. The information about the stacking is encoded in the $x$, $y$ cut taken at $z=\frac{1}{3}=5.73$~\AA{}, which is the $z$ value equal to the distance between adjacent layers in the crystal. Interpreting this segment of the 3D-$\Delta$PDF could, in principle, be done by constructing and fitting a model, but this would be rather complicated, as it would contain many inter-atom vectors; however, we immediately observe a similarity between the $z=\frac{1}{3}$ cuts shown in Figure~\ref{fig2}(c)\&(d) and the $z=0$ cuts shown in Fig.~\ref{fig2}(a)\&(b) (at 300 and 30~K respectively). At 300~K the $z=\frac{1}{3}$ cut resembles the $z=0$ cut, but shifted by an in-plane vector of $(\frac{2}{3},0)$ which is the stacking type in the long range $C2/m$ crystal structure. Indeed, if the disorder in RuCl$_3$ is entirely composed of stacking faults without any in-plane disorder, the $z=\frac{1}{3}$ 3D-$\Delta$PDF should theoretically be equal to a weighted linear sum of $z=0$ 3D-$\Delta$PDF data shifted in-plane according to the stacking types. 

We therefore fit the $z=\frac{1}{3}$ $\Delta$PDF using a model consisting of a weighted combination of shifted $z=0$ $\Delta$PDFs, with the weighting factors as free parameters. Since all the different shifts applied lead to density at different positions in the $z=\frac{1}{3}$ $\Delta$PDF, there is one unique best fit solution, with weighting factors that are independent of one another. The result of this fit is shown in Figure~\ref{fig2}(e)\&(f), and shows that our model provides a good representation of the $z=\frac{1}{3}$ plane. The in-plane translation vectors used to generate these linear sums are shown in Fig.~\ref{fig2}(g)\&(h), which also indicate the percentage contribution of each translation vector. The presence of these types of $(\frac{1}{3}, 0)$ and $(0, \frac{1}{3})$ stacking faults in RuCl$_3$ is understandable since they maintain a close packed structure for the Cl atoms of adjacent layers, and so these are the minimum energy stacking faults. Errorbars on the weighting factors were estimated by manually adjusting the weighting factors until the weighted sum differed visibly from the $z=\frac{1}{3}$ data. Errors on the calculated percentage contributions were computed by summing in quadrature the contributions from the weighting factors.

In the 3D-$\Delta$PDFs at both 30~K and 300~K we observe that the dominant stacking is that of the long range ordered phase ($R\bar{3}$ and $C2/m$ for 30~K and 300~K respectively). This is manifested in the raw data as the presence of extended tails from the very intense main Bragg peaks [Fig.~\ref{fig1}(c)], which contribute to the diffuse signal as these tails are larger than the punch size used to compute the 3D-$\Delta$PDF. These stacking types are indicated by the solid green arrows in the right-hand panels of Fig.~\ref{fig2}. RuCl$_3$ crystals often contain contributions from three possible $C2/m$ domains, rotated by $120^{\circ}$ about the $c$ axis. However the crystal selected for this measurement was chosen for its low degree of twinning, and so one of the three possible monoclinic domains is predominant over the other two at 300~K. In the low temperature phase, the 3D-$\Delta$PDF is consistent with an equal proportion of the two possible rhombohedral-like domains, as is reflected in Fig.~\ref{fig2}(g)\&(h). We note that the percentage contribution of this long range ordered phase varies somewhat depending on the punch size chosen, however the relative contributions of the different stacking faults (described below) do not depend on punch size. This was tested by computing the $\Delta$PDFs with two different punch sizes (radius $0.12$ and $0.16$ reciprocal lattice units (RLU)). The relative contributions of the stacking faults were very similar for these two choices of punch size. Figure~\ref{fig2} was generated with a punch size of $0.16$ RLU which has a somewhat reduced contribution from the long range ordered stacking.

The RuCl$_3$ stacking faults that differ from those corresponding to the nominal long range ordered phase are shown by the blue, solid black, and red arrows of Fig.~\ref{fig2}(g)\&(h). The predominant type of stacking fault in the 300~K phase is from the secondary monoclinic twin domains, despite the fact that the main Bragg peaks do not indicate long range ordered twin domains. In both high and low temperature phases, we also observe an intermixing with the stacking type found in the other structural phase. Namely, $C2/m$ type stacking in the low temperature phase and $R\bar{3}$ type stacking in the high temperature phase. We observe that although the long range ordered $C2/m$ structure is primarily untwinned, the three types of $C2/m$ stacking faults are observed in equal proportions in the low temperature crystal. In the high temperature phase, we also infer the presence of a direct stacking of one layer on top of another indicated by the red arrow at zero shift, which is required to obtain a good match between the $z=\frac{1}{3}$ data and the weighted sums (see Fig.~\ref{fig2}). This type of stacking is not observed in either $C2/m$ or rhombohedral-like long range order.

\section{Intra-layer order}
Information about possible distortions in the single layer local structure can also be extracted from our data. This is interesting because the structure of a single layer of RuCl$_3$ exfoliated onto a silicon wafer proved to be substantially different from the structure of the nominal space groups \cite{Yang2023}, suggesting that RuCl$_3$ may have a propensity towards local distortions. It is also notable that theoretical calculations usually assume the undistorted room temperature structure. We show here that there is no evidence of such a distortion in either the long range ordered low temperature phase, or in the 3D-$\Delta$PDFs computed from the diffuse scattering. We cannot, however, exclude the possibility of distortions smaller than our 0.3~\AA{} resolution.

Panels (a)\&(b) of Fig.~\ref{fig3} show the 3D-$\Delta$PDF corresponding to the structure within the planes. This is taken at a small value of $z=0.08$ which corresponds to the Ru-Cl distance. Such a distance was chosen over $z=0$ because this plane is more sensitive to the distortions observed in Ref.~\cite{Yang2023}. To test whether this short-range information in the 3D-$\Delta$PDF is consistent with the measured average long range structure, we created a ``pseudo $\Delta$PDF'' that reflects the long range structure in a way that can be easily compared to the short range disorder in the $\Delta$PDF. We first computed the Patterson function by isolating and Fourier transforming only the Bragg peaks. The $z=0.08$ cut of the Patterson function cannot be directly compared to the 3D-$\Delta$PDF computed from the diffuse scattering, and instead needs to be converted to a difference Patterson function compared to the unit cell with in-plane lattice parameters $\frac{a}{3}$ (described in the previous section). The average Patterson function for this reduced unit cell was computed by first folding and summing the $(x=6, y=3, z=0.08)$ cut onto a 2D array $\frac{1}{3} \times \frac{1}{3}$ the original size. This small array was tiled $3\times3$ to generate the average cut for the reduced unit cell. This average was subtracted from the data to generate the ``pseudo $\Delta$PDF'' that can be compared to the diffuse scattering $\Delta$PDF.

The net result of the ``pseudo $\Delta$PDF'' analysis is shown in (c)\&(d) of Fig.~\ref{fig3} and appears similar to panels (a)\&(b). This suggests that, within our resolution, there is no evidence for short range distortions within the layers. In performing this comparison we note that our upper momentum limit of the data collected gives us a somewhat crude real space resolution of $\sim$0.3~\AA{}, distortions smaller than this would not be detected. Both data were also found to have a pattern that is similar to the nominal structure by simulating the ``pseudo $\Delta$PDF'' for the $C2/m$ space group in an equivalent way. The rather large distortions on the order of those seen in a single exfoliated layer would have been readily visible. As reported in Ref.~\cite{Yang2023}, in a single exfoliated layer the flat Ru layers become buckled, the Ru-Cl distance decreases and the Cl layers show an in-plane distortion. A comparison between the data in Fig.~\ref{fig3}(a) to a calculation in Fig.~\ref{fig3}(f) which computes the 3D-$\Delta$PDF based on the structure reported in \cite{Yang2023} makes it clear that the two structures are very different and that no distortions of this magnitude occur in the structural transition of the bulk crystal.

\section{Conclusions}

Our x-ray diffraction data on a single crystal of $\alpha$-RuCl$_3$ shows a dramatic structural change from a single domain $C2/m$ structure at room temperature to a completely twinned rhombohedral-like structure at 30~K. Although the diffraction pattern at low temperature could be well indexed by the  $R\bar{3}$ unit cell previously proposed for this material, the crystal does not show three-fold symmetry but rather ``remembers'' its single domain $C2/m$ structure upon warming above the transition temperature suggesting it is not truly three-fold symmetric. 

Temperature cycling does, however, introduce disorder into the crystal which manifests as rods of diffuse scattering perpendicular to the honeycomb planes. This pattern of diffuse scattering is typical of mis-stacked or translated honeycomb layers, and we evaluated the types of mis-stackings via 3D-$\Delta$PDF. We find that the stacking disorder is primarily an intermixing of the high and low temperature structures, combined with rotational twinning. The low temperature phase showed an equal mixture of all three $C2/m$ stacking types, despite beginning with an untwinned $C2/m$ crystal at room temperature. At room temperature we also observed stacking with no in-plane translation between adjacent layers, unlike that seen in either long-range ordered structure. Our measurements therefore give a comprehensive overview of how the three dimensional crystal structure of $\alpha$-RuCl$_3$ is built up from the component honeycomb layers. 

Stacking disorder is thought to be at the root of sample variability seen in both magnetic and transport properties of RuCl$_3$ \cite{banerjee2016, zhang2023sampledependent, zhang2023stacking}, however the mechanisms behind these effects are not fully understood. Introducing stacking faults by physically deforming the crystal is known to nearly double the magnetic ordering temperature \cite{banerjee2016}. Variability is also seen in thermal Hall measurements, with only some samples showing the half integer quantization characteristic of the Majorana edge modes expected in a Kitaev quantum spin liquid \cite{zhang2023sampledependent, zhang2023stacking}. It has been shown that the observability of the quantized thermal Hall effect is related to the coupling of these particles with the phonon bath \cite{Ye2018}. Furthermore, phonons with out-of-plane atomic motions or propagation vectors are now known to play a particularly important role \cite{mu2022}. Since out-of-plane phonons are strongly affected by the layer stacking, stacking faults may therefore determine whether the quantized thermal Hall effect is observed.

Despite the evident importance of the layer stacking, RuCl$_3$ is often considered in the two-dimensional limit, without examining the effects of interactions between the honeycomb layers. Our measurement shows how the layers are stacked, and therefore provides a crucial basis for understanding the role of the third dimension. The stacking will determine the distances between ruthenium atoms in adjacent layers, affecting the interlayer magnetic interactions and magnetic ordering. Stacking faults will likely also affect the phonon dispersions important for the thermal Hall effect measurements. Density functional theory calculations have indeed been reported using ordered structures with different stacking types, showing substantial changes due to stacking \cite{mu2022}. The effects of stacking faults could be emulated using larger supercell structures containing stacking faults of the types identified in our work, although we note that such simulations will be computationally expensive. By identifying the types of layer stacking that occur in a single crystal, our measurement therefore provides an important basis for understanding the effects of disorder on the remarkable electronic and magnetic behavior of RuCl$_3$.

\begin{acknowledgments}
Work at Brookhaven is supported by the Office of Basic Energy Sciences, Materials Sciences and Engineering Division, U.S.\ Department of Energy (DOE) under Contract No.\ DE-SC0012704. Work at Argonne National Laboratory (MJK, RO, SR, single crystal diffuse scattering measurements and data reduction) was supported by the U.S.\ Department of Energy, Office of Science, Basic Energy Sciences, Materials Sciences and Engineering Division. Work at Oak Ridge National Laboratory was sponsored by the U.S. Department of Energy, Office of Science, Basic Energy Sciences, Materials Sciences and Engineering Division. This research used resources of the Advanced Photon Source, a U.S.\ Department of Energy (DOE) Office of Science User Facility operated for the DOE Office of Science by Argonne National Laboratory under Contract No.~DE-AC02-06CH11357. Crystal synthesis work by JQY was supported by the U.S.\ Department of Energy, Office of Science, National Quantum Information Science Research Centers, Quantum Science Center. Crystal structures were plotted using VESTA \cite{vesta}.
\end{acknowledgments}

\bibliography{refs}

\end{document}